\documentclass[aps,pre,twocolumn,longbibliography,preprintnumbers,showpacs,amsmath,amssymb,amsfonts,nofootinbib]{revtex4-2}
\usepackage{graphicx}
\usepackage{dcolumn}
\usepackage{bm}
\usepackage[usenames,dvipsnames]{color}
\usepackage{multirow}
\usepackage{soul}
\usepackage{natbib}
\usepackage{amsmath, amssymb}
\usepackage{bbold}
\usepackage{physics}
\usepackage{cancel}
\usepackage[hypertexnames=false]{hyperref}

\definecolor{magenta}{rgb}{0.8,0.2,0.8}

\begin{document}

\title{Influence of coupling symmetries and noise on the critical dynamics of synchronizing oscillator lattices}

\author{Ricardo Guti\'errez }
\author{Rodolfo Cuerno}
\affiliation{Grupo Interdisciplinar de Sistemas Complejos (GISC), Departamento de Matemáticas, Universidad Carlos III de Madrid, 28911 Legan{\'e}s, Madrid, Spain}

\preprint{Physica D {\bf 473}, 134552 (2025)}

%% Abstract
\begin{abstract}
%% Text of abstract
Recent work has shown that the synchronization process in lattices of self-sustained (phase and limit-cycle) oscillators displays universal scale-invariant behavior previously studied in the physics of surface kinetic roughening. The type of dynamic scaling ansatz which is verified depends on the randomness that occurs in the system, whether it is columnar disorder (quenched noise given by the random assignment of natural frequencies), leading to anomalous scaling, or else time-dependent noise, inducing the more standard Family-Vicsek dynamic scaling ansatz, as in equilibrium critical dynamics. The specific universality class also depends on the coupling function: for a sine function (as in the celebrated Kuramoto model) the critical behavior is that of the Edwards-Wilkinson equation for the corresponding type of randomness, with Gaussian fluctuations around the average growth. In all the other cases investigated, Tracy-Widom fluctuations ensue, associated with the celebrated Kardar-Parisi-Zhang equation for rough interfaces.
%the nonequilibrium universality classes observed are those of growth processes in the presence of columnar disorder, which display anomalous scaling. In contrast, in the presence time-dependent noise, they are those of growth processes subject to thermal noise, following the standard Family-Vicsek dynamic scaling ansatz familiar from equilibrium critical dynamics. Moreover, if the coupling function is a sine function (as in the celebrated Kuramoto model), the critical behavior is that of the Edwards-Wilkinson equation subject to quenched or time-dependent noise, with Gaussian fluctuations around the average growth. In all the other cases investigated, Tracy-Widom fluctuations ensue, which are associated with the celebrated Kardar-Parisi-Zhang equation for a rough interface far from equilibrium.
Two questions remain to be addressed in order to complete the picture, however: 1) Is the atypical scaling displayed by the sine coupling preserved if other coupling functions satisfying the same (odd) symmetry are employed (as suggested by continuum approximations and symmetry arguments)? and 2) how does the competition between both types of randomness (which are expected to coexist in experimental settings) affect the nonequilibrium behavior? We address the latter question by numerically characterizing the crossover between thermal-noise and columnar-disorder criticality, and the former by providing evidence confirming that it is the symmetry of the coupling function that sets apart the sine coupling, among other odd-symmetric couplings, due to the absence of Kardar-Parisi-Zhang fluctuations.
\end{abstract}

\maketitle

\section{Introduction}

Synchronization is known to be relevant in many complex dynamical phenomena in a myriad of systems, ranging from electronic circuits and coupled qubits to cortical columns in the brain, including even manifestations of collective animal and human behavior \cite{pikovsky, boccaletti_book}. As such, it is perhaps not surprising that is has become one of the main research topics in nonlinear physics, both theoretically and experimentally, in few- and many-body systems of oscillators, including lattices but also networks with complex coupling topologies \cite{osipov,arenas}. The oscillators themselves may be quantum or classical, regular or chaotic, as indeed different notions of synchronization have been proposed to characterize synchronous motion in such a diversity of systems \cite{pikovsky,boccaletti}. 

Despite this great variety, a basic form of competition lies invariably at the heart of most synchronization studies: there is some kind of coupling between the oscillating units that tends to make their dynamics more similar, and a source of randomness that makes them different, whose opposite effect must be overcome by the coupling in order to reach synchronization. Quite naturally, the most commonly addressed questions when facing such scenarios are: {\it Under which conditions (e.g., for which couplings relative to the strength of the noise) does the system synchronize?} and {\it What kind of transition to synchronization is observed as the coupling (or some other control parameter) is varied?} In essence, they both deal with a static (steady-state) picture of synchronous dynamics.

A much less studied question is the following: {\it Given the right conditions for synchronization, what kind of dynamical process leads from an arbitrary initial state to a synchronous evolution at long times?} Such a relevant dynamical question has received surprisingly little attention, probably due to the implicit assumption that the answer must be highly dependent on the specific dynamical systems that synchronize, and even their parameter values. The expectation that this is a sensitively system-dependent process may also explain why a mathematical connection between the synchronization dynamics and celebrated models of surface growth \cite{barabasi,krug97}, most notably the Kardar-Parisi-Zhang (KPZ) equation \cite{takeuchi}, has been repeatedly mentioned in the past (see, e.g.\!, \cite{kuramoto_book,pikovsky}), but had not been fully explored, in spite of some recent works that do indeed exploit relevant aspects thereof \cite{moroney,lauter}. 

Against this background, our recent work \cite{PRR1,PRR2,PRE} has provided a first (to our knowledge) thorough investigation of the connection between synchronization and surface kinetic roughening. The analysis is based on studying the ``phase interface'' associated with a ring of oscillators by means of numerical simulations, continuum approximations, and phase-reduction methods, and it focuses on observables and concepts employed in the study of growing rough surfaces and interfaces \cite{barabasi, krug97}. The models considered so far \cite{PRR1,PRR2,PRE} are one-dimensional systems of phase oscillators of the Kuramoto-Sakaguchi (KS) type \cite{sakaguchi}, and limit-cycle (van der Pol \cite{strogatzbook} and Stuart-Landau \cite{introsl}) oscillators. This work encompasses the two most widely studied types of randomness in synchronization research, namely, a random assignment of natural frequencies \cite{PRR1,PRR2}, which we call columnar disorder, as that form of quenched disorder is referred to in surface kinetic roughening \cite{halpinhealy}, and time-dependent noise, see Ref.\ \cite{PRE}. 

The main findings in these recent references can be summarized as follows: for large space-time scales the dynamics of the synchronization process is robust against variations in the dynamical units and their parameters, and generically displays the universal Tracy-Widom (TW) distribution for fluctuations around the average growth of the KPZ equation with time-dependent noise \cite{takeuchi}. The critical exponent values, as well as the covariance, depend on the nature of the randomness, being those of the columnar KPZ equation for systems with columnar disorder and those of the standard KPZ equation with time-dependent noise. 
%the universal behavior of the KPZ equation with the type of randomness considered in the microscopic oscillator lattice model (either columnar disorder or thermal noise). This includes not only the critical exponent values, but also the universal Tracy-Widom (TW) fluctuations around the average growth and other universal features. (In fact, TW fluctuations are part of the 1D KPZ universality class in the presence of thermal noise \cite{takeuchi}, but as far as we are aware it is not yet know if that is also true for the universal behavior of the KPZ with columnar disorder.) 
An exception to this generic behavior is known to exist, however, when the oscillator phases couple through a sine function, as in the Kuramoto model \cite{kuramoto_book,acebron}: in that case the universal features are those of the Edwards-Wilkinson (EW) equation (i.e.\! the linearized KPZ equation \cite{barabasi,krug97}) with the corresponding type of randomness, including the critical exponent values and the (Gaussian) fluctuations. The covariance, however, is not changed by this particular choice of the coupling function, as it does not seem to be affected by the KPZ nonlinearity for either type of randomness (in the case of time-dependent noise, that is expected as it is known to happen for the corresponding universality classes \cite{takeuchi}). Both KPZ and EW universality, for either type of randomness, are examples of generic scale invariance (GSI), i.e., scale invariance that occurs generically, meaning that its observation does not require setting parameters to specific (critical) values or other zero-measure sets in parameter space, unlike the typical situation in equilibrium critical dynamics \cite{tauber14}.

In this work, we numerically explore one-dimensional systems of phase oscillators similar to those considered in Refs.~\cite{PRR1,PRE} in order to address two relevant questions that have not yet been addressed in those previous contributions: 1) Does the exceptional large-scale (EW) behavior found for the Kuramoto (sine) coupling function persist for other coupling functions with the same symmetry? 2) What happens when both types of randomness (columnar disorder and time-dependent noise) are simultaneously present (as is likely the case in most experimental settings)?  While addressing the latter serves to highlight the dominant role of columnar disorder, and to charaterize a crossover between time-dependent-noise (at earlier times) and columnar-disorder (at later times) universal behavior, the former will allow us to conclude that it is the odd symmetry of the coupling function (and not a specific sine coupling term) that leads to EW universality. There were already some indications that this might be the case, including the up-down symmetries of the oscillator system and its continuum approximation \cite{PRR1}, and prior results related to the threshold for synchronization and its scaling with the system size \cite{strogatz,ostborn}; see  Ref.~\cite{PRR1} for more details.

The structure of the paper is as follows. In Section 2 we describe the phase-oscillator model under consideration (which is an extension of the KS model including higher harmonics, time-dependent noise and columnar disorder) and its continuum approximation, paying special attention to the role played by the odd symmetry of the coupling function,  which later on provides a guide to understanding the large-scale behavior found in the numerical results. Section 3 includes a description of the kinetic-roughening observables considered in the analysis of the simulation data of subsequent sections, and also the scaling forms involved. In Section 4, we numerically investigate the dependence of the universal GSI behavior on the symmetry of the coupling function, for both time-dependent noise and columnar disorder (yet without combining them). In Section 5 we study the impact of the competition between these two forms of randomness, which are now simultaneously present in our simulations, on the GSI behavior at large scales. Finally, there is a concluding section where we summarize our findings, mention some open questions that remain to be addressed in future work, and highlight the main consequences of our results for the experimental investigation of GSI in the synchronization of oscillating lattices and media, a research topic yet to be initiated.

\section{Phase-oscillator lattice model and continuum approximation}

\subsection{General model of an oscillator lattice}

We consider a system of phase oscillators (i.e., idealized dissipative dynamical systems with an attracting limit cycle) on a $d$-dimensional hypercubic lattice of linear size $L$. The state of each oscillator is given by a phase $\phi_i(t)$, which evolves in time autonomously with a natural frequency $\omega_i$, interacts diffusively with its neighbors through a coupling function $\Gamma(\cdot)$, and is subjected to external noise $\xi_i(t)$:
\begin{equation}
\frac{d \phi_i(t)}{d t} = \omega_i + \sum_{j\in \Lambda_i} \Gamma[\phi_j(t) -\phi_i(t)] + \xi_i(t),
\label{eq1}
\end{equation}
for $i=1,2,\ldots, L^d$, where $\Lambda_i$ is the set of $2 d$ neighbors of site $i$. This is a very general setting for coupled phase oscillators on a lattice, which does not specify the number of spatial dimensions $d$ or the form of the coupling function $\Gamma(\cdot)$, and may contain both columnar disorder (if the natural frequencies $\omega_i$ are randomly assigned) and time-dependent noise $\xi_i(t)$. Synchronization is said to take place whenever the coupling overcomes the effect of randomness and all the oscillators attain the same effective frequency of oscillation, defined as
\begin{equation}
\omega^\text{eff}_i \equiv \lim_{T\to\infty}\frac{\phi_i(\tau+T)-\phi_i(\tau)}{T},
\label{omegaeff}
\end{equation}
where $[0,\tau]$ is a time interval sufficiently long to contain the transient behavior, and the limit is assumed to exist. This relatively weak form of synchronization is frequently referred to as frequency locking or frequency entrainment in the literature.

\subsection{Continuum approximation}
\label{sec:contapp}

In our simulations below we will focus on specific instances of Eq.~(\ref{eq1}), but before particularizing it we shall briefly analyze its large-scale properties through a continuum approximation previously developed in Refs.~\cite{PRR1,PRE}. Generalizing the approach of Ref.~\cite{kuramoto_book}, we write the positions of the oscillators as vectors in continuous space, ${\bf x} = (x_1, \ldots, x_d)\in \mathbb{R}^d$, so the phase of oscillator $i$, $\phi_i$, is denoted $\phi({\bf x})$, and the neighboring oscillators are placed at positions  ${\bf x}\pm a {\bf e}_k$, where ${\bf e}_k$ is a canonical basis vector, $a$ is the lattice constant, and $k=1,\ldots, d$. Thus Eq.~(\ref{eq1}) becomes
\begin{align}
\partial_t \phi({\bf x},t) = \eta({\bf x},t) &+ \sum_{k=1}^d \left(\Gamma[\phi({\bf x}+ a {\bf e}_k,t) -\phi({\bf x},t)]\right.\nonumber\\
&\left.+ \Gamma[\phi({\bf x}- a {\bf e}_k,t)  -\phi({\bf x},t)]\right),
\label{eq2}
\end{align}
where $\eta({\bf x},t)$ represents the combined effects of columnar disorder and time-dependent noise, including possible (expected) renormalizations of parameters. In a coarse-grained description, where the perturbative parameter $a$ is assumed to be small compared to the wavelengths over which $\phi({\bf x},t)$ fluctuates, a Taylor expansion of the phase field $\phi({\bf x})$ around ${\bf x}$, and of the coupling function $\Gamma(\cdot)$ around the origin, yields
%\begin{widetext}
\begin{align}
\partial_t \phi ({\bf x},t)  &=\eta^*({\bf x},t) 
+ a^2 \Gamma^{(1)}(0)\!\sum_{k=1}^d\!\partial_k^2 \phi({\bf x},t) \nonumber\\
&+  a^2 \Gamma^{(2)}(0)\!\sum_{k=1}^d\!(\partial_k \phi({\bf x},t) )^2 +  \mathcal{O}(a^4),
%+ a^2 \Gamma^{(1)}(0)\!\sum_{k\in \mathbb{N}_d}\!\partial_k^2 \phi({\bf x},t) \nonumber\\
%&+  a^2 \Gamma^{(2)}(0)\!\sum_{k\in \mathbb{N}_d}\!(\partial_k \phi({\bf x},t) )^2 +  \mathcal{O}(a^4),
%+ \frac{a^4}{12}\!\left[\Gamma^{(1)}(0)\!\sum_{k\in \mathbb{N}_d}\!\partial_k^4 \phi \right.\nonumber\\
%&\left. + \Gamma^{(2)}(0)\!\left(3\!\sum_{k\in \mathbb{N}_d}\!(\partial_k^2 \phi)^2 + 4\!\sum_{k\in \mathbb{N}_d}\!\partial_k \phi\, \partial_k^3 \phi \right)+ 6 \Gamma^{(3)}(0)\!\sum_{k\in \mathbb{N}_d}\!(\partial_k \phi)^2 \partial_k^2 \phi+ \Gamma^{(4)}(0)\!\sum_{k\in \mathbb{N}_d}\!(\partial_k \phi)^4 \right]+   \mathcal{O}(a^6),
\label{eq3}
\end{align}
%\end{widetext}
where $\Gamma^{(n)}(\cdot)$ denotes the $n$-th derivative of $\Gamma(\cdot)$. The asterisk in $\eta^*({\bf x},t)$ simply indicates that a term proportional to $\Gamma(0)$ has been absorbed into the randomness. This can be conveniently removed by studying oscillators in a co-moving frame, as typically done in the study of synchronization \cite{acebron}.%, and also in interfaces with constant and spatially uniform driving terms \cite{barabasi}.

For a relatively slow spatial variation of the phase field $\phi({\bf x},t)$, as occurs for couplings well into the synchronized regime, it may be reasonable to neglect terms of order higher than $a^2$, which yields the effective continuum equation
\begin{equation}
\partial_t \phi({\bf x},t) =   \eta^*({\bf x},t) + \nu \nabla^2 \phi({\bf x},t) + \frac{\lambda}{2} [\nabla \phi({\bf x},t)]^2,
\label{eq4}
\end{equation}
where as usual the Laplacian $\nabla^2 \phi({\bf x},t) \equiv \sum_{k=1}^d\!\partial_k^2 \phi({\bf x},t)$ and the squared norm of the gradient  $[\nabla \phi({\bf x},t)]^2\equiv \sum_{k=1}^d\![\partial_k \phi({\bf x},t)]^2$. We have introduced two parameters, namely $\nu \equiv  a^2 \Gamma^{(1)}(0)$ and $\lambda/2 \equiv  a^2 \Gamma^{(2)}(0)$, following the standard notation in the surface growth literature, where they quantify the surface tension and interface growth along the local surface normal direction (sometimes referred to as the KPZ nonlinearity), respectively \cite{barabasi,krug97}. A situation that will be considered later is that in which $\lambda = 0$. Specifically, we will focus on cases where $\Gamma(\cdot)$ is an odd function, which implies that its second derivative is too, %(use the fact that the derivative of an even/odd function is odd/even twice)
hence %vanishes at the origin, 
$\Gamma^{(2)}(0) = 0$. Studying this question is one of the two aims of this work. 

Despite the uncontrolled approximations introduced, this derivation has proven relevant to explain the large-scale behavior of the synchronization process as observed in numerical results based on both phase and limit-cycle oscillators, in the presence of columnar disorder or time-dependent noise \cite{PRR1,PRR2,PRE}. In this regard, Eq.~\eqref{eq4} features the same deterministic derivative terms as the KPZ equation \cite{kardar}, and numerical analyses based on lattices of oscillators indeed show that their synchronization process has robust universal features pertaining to the KPZ equation with columnar disorder \cite{PRR1,PRR2} or time-dependent (thermal) noise \cite{PRE}, depending on the type of randomness introduced in the microscopic model. Note that the KPZ equation with columnar disorder and its linear version for $\lambda = 0$, the EW equation with columnar disorder (also known as the Larkin model \cite{purrello}), while much less studied than the time-dependent-noise KPZ equation, and its linearized version, the time-dependent noise EW equation \cite{barabasi,krug97,takeuchi}, have also been the focus of some attention in the literature, and are known to possess distinct features very different from those of the  time-dependent-noise universality classes \cite{halpinhealy, szendro, purrello}. The emerging large-scale behavior that results from the competition between both forms of randomness, i.e., columnar disorder and time-dependent noise, when simultaneously present (as in almost any conceivable experimental setting), has not yet been addressed in a synchronization context as far as we are aware. Studying this competition is in fact the second main aim of this work.

\subsection{One-dimensional oscillator lattices with multi-harmonic couplings}
\label{sec:1dmodels}

In our simulations below, we focus on a one dimensional ($d=1$) version of Eq.~(\ref{eq1}). Specifically, the $L$ oscillators will always be at the sites of a ring, i.e., a chain with periodic boundary conditions (PBCs). Moreover, as the coupling function $\Gamma(\cdot)$ is generally assumed to be $2\pi$-periodic and smooth, for simplicity we focus on couplings that can be described with just a few Fourier harmonics:
\begin{equation}
\Gamma[\Delta \phi(t)] = K \sum_{n=1}^p \left(a_n \cos[n \Delta \phi(t)] + b_n \sin[n \Delta \phi(t)]\right),
\label{gamma}
\end{equation}
where $\Delta \phi(t)$ will be, for a given oscillator $i$, either $\phi_{i+1}(t)-\phi_i(t)$ or $\phi_{i-1}(t)-\phi_i(t)$, and $p$ is the number of harmonics considered. While $K$ is the coupling strength, the coefficients $a_i$ and $b_i$ for $i=1,2,\ldots, p$ are chosen so that the normalization condition $K^{-2} \int_{-\pi}^\pi \Gamma^2(\Delta \phi) d(\Delta \phi) = \pi \sum_{n=1}^p \left(a_n^2 + b_n^2\right) = 1$ is satisfied. This choice places the overall strength of the coupling on $K$, irrespective of the number and relative weights of the different Fourier components. Moreover, we only study attracting couplings, for which $\Gamma^{(1)}(0) = b_1 + 2\, b_2 + \cdots + p\, b_p > 0$, which in the continuum approximation above, Eq.\ (\ref{eq4}), amounts to having a positive surface tension $\nu$. In the particular case when $p=1$, the coupling function is sometimes written in the KS form \cite{sakaguchi,sakaguchi86}, $\Gamma(\Delta \phi) \propto \sin(\Delta \phi + \delta)$, with $\delta \in (-\pi/2,\pi/2)$, see e.g.\! \cite{PRR1,PRE}, where the Fourier coefficients in Eq.~(\ref{gamma}) are $a_1 = (\sin\delta)/\sqrt{\pi}$ and $b_1 = (\cos\delta)/\sqrt{\pi}$, with $\tan \delta$ thus giving the relative strength of $\cos \Delta \phi$ to $\sin \Delta \phi$.

One property that has been highlighted in the literature is the symmetry of the coupling function under phase inversion, $\Delta \phi \to -\Delta \phi$, namely, whether the function is odd, $\Gamma(\Delta \phi) + \Gamma(-\Delta \phi) = 0$, or not \cite{strogatz, ostborn}. This symmetry has crucial implications for several large-scale dynamical features of synchronization, which can be related (as mentioned above) to the occurrence of the nonlinearity in the continuum approximation given by Eq.~(\ref{eq4}), i.e.\! whether $\lambda \neq 0$. An obvious feature already in the oscillator model, Eq.\ \eqref{eq1}, is that only when $\Gamma(\Delta\phi)$ is odd does the system have up-down symmetry, i.e., invariance under phase reversal $\phi_i\to -\phi_i$, in a statistical sense, provided that the columnar disorder and time-dependent noise distributions are (evenly) symmetric around their means \cite{PRR1}. In the case of harmonic ($p=1$) KS coupling, the odd symmetry only holds for $\delta = 0$, which corresponds to $\Gamma(\Delta \phi) \propto \sin(\Delta \phi)$, as in the Kuramoto model \cite{acebron}. For this coupling form, the large-scale dynamics of the synchronization process has indeed been shown to be in the universality class of the EW equation with columnar disorder \cite{PRR1} or time-dependent noise \cite{PRE}, as the case may be.

We consider the possibility of having both columnar disorder and time-dependent noise at the same time  in our model, as both forms of randomness will appear in the simulations below, sometimes separately, sometimes simultaneously. In the case of columnar disorder, the natural frequencies are independent and identically distributed according to a Gaussian with zero mean and standard deviation $\sigma_\text{col}$, i.e., $\langle \omega_i \rangle = 0$  and $\langle \omega_i \omega_j \rangle = \sigma_\text{col}^2\, \delta_{i j}$, where $\delta_{i j}$ is the Kronecker delta. Concerning the time-dependent noise terms, they are independent and Gaussian-distributed as well, with zero mean and standard deviation $\sigma_\text{tdep}$, and delta-correlated in time, $\langle \xi_i(t) \rangle = 0$, $\langle \xi_i (t)\, \xi_j (t')\rangle = \sigma_\text{tdep}^2\, \delta_{ij}\, \delta(t-t')$, where $\delta(\cdot)$ is the Dirac delta.

\subsection{Numerical simulations}
\label{numsim}

In our simulations, rings of $L=1000$ phase oscillators with coupling $\Gamma(\Delta \phi)$ as in Eq.~(\ref{gamma}) with $p\leq 3$ harmonics are considered. Specifically, the following five coupling functions are studied:
\begin{itemize}
\item Kuramoto coupling (K), for which $b_1 \neq 0$, the remaining coefficients being zero (this can also be thought of as KS coupling with $\delta = 0$);
\item Kuramoto coupling with higher-order sines (KH1), for which $b_1 \neq 0$, $b_2 = b_1/2$, $b_3 = b_1/4$, and $a_n = 0$  for $n=1,2,3$;
\item Another version of Kuramoto coupling with higher-order sines (KH2), for which $b_1 \neq 0$, $b_2 = -b_1/2$, $b_3 = b_1/4$, and $a_n = 0$  for $n=1,2,3$;
\item KS coupling with $\delta = \arctan 5$ (which we will simply refer to as KS), for which $b_1 \neq 0$, $a_1= 5 b_1$, the remaining coefficients being zero;
\item Cosine coupling with higher-order terms (CH), for which $a_1 \neq 0$, $a_2 = a_1/2$, $a_3 = a_1/4$, $b_3 = a_1/10$ (a small sine term to make the coupling attracting), and $b_1 = b_2 = 0$.
\end{itemize}
In every case, we specify whether the Fourier coefficients are nonzero and, if so, their relative magnitudes, their specific values being determined by the normalization condition mentioned above.  The crucial aspect of this coupling functions in this context is that K, KH1, and KH2 are odd coupling functions for which $\Gamma(\Delta \phi)+\Gamma(-\Delta \phi) = 0$, while KS and CH do not satisfy that condition due to the presence of cosine terms. As mentioned above, K has been previously studied both in the presence of columnar disorder \cite{PRR1} and time-dependent noise \cite{PRE}, in both cases yielding the behavior of the EW equation with the corresponding type of randomnness. KS with time-dependent noise, on the other hand, is one of the cases explored in Ref.~\cite{PRE}, and this and similar couplings (with other nonzero values of $\delta$) have been studied in the presence of columnar disorder in Ref.~\cite{PRR1}, showing the large-scale behavior of the columnar KPZ equation (again for the corresponding type of randomness). The reason why we consider three cases where the (unusual, apparently requiring an exact symmetry in the coupling) EW behavior is observed and just two where the (typical) KPZ is found is that the generic appearance of KPZ scaling  when the odd symmetry of the coupling function is not present has already been well established, even in the case of limit-cycle oscillators \cite{PRR2,PRE}, whereas EW scaling has only been found for K thus far \cite{PRR1,PRE}. In this regard, we could just consider one non-odd coupling function in our analysis, but we include two (KS and CH) in order to illustrate the existence of EW-to-KPZ crossover effects to be discussed below.

The overall coupling strength is taken to be $K=20$, found to be above the threshold for synchronization for all couplings and forms of randomness. All rings of oscillators start from a flat initial condition, $\phi_i(0)=0$ for all $i$, and their
time evolution is numerically unravelled with a time step $\delta t = 0.01$. To this end, a fourth-order Runge-Kutta algorithm is employed in the deterministic situation of zero time-dependent noise (only columnar disorder, as in Sec.\! \ref{col}), while an Euler-Mayurama scheme \cite{Toral} is used for time-dependent noise, as in Sec.\! \ref{therm}, or when both types of randomness are simultaneously present, as in Sec.\! \ref{comp}. 

Concerning the randomness, when time-dependent noise is considered on its own, as in Sec.\! \ref{therm}, it has an intensity (standard deviation)  $\sigma_\text{tdep} = 0.1$ (with $\sigma_\text{col} = 0$). And the columnar disorder, in its turn, when considered as the only source of randomness, as in Sec.\! \ref{col}, also has an intensity $\sigma_\text{col} = 0.1$ (while $\sigma_\text{tdep} = 0$ in that case). In Sec.\! \ref{comp}, where the competition between the two types of randomness is investigated, different (nonzero) values of $\sigma_\text{col}$ and $\sigma_\text{tdep}$ are inspected, namely $\sigma_\text{col} = 0.1$ and $\sigma_\text{tdep} = 0.01$, $\sigma_\text{col} = 0.01$ or $0.02$ (depending on the coupling function) and $\sigma_\text{tdep} = 0.1$, and $\sigma_\text{col} = 0.0001$ and $\sigma_\text{tdep} = 0.1$. The latter choices allow us to explore different situations ranging from dominant columnar noise to dominant time-dependent noise, passing through a crossover where the time-dependent noise prevails at the initial stages, to be  replaced by columnar-disorder scaling later on.

%\section{Morphological analysis and generic scale invariance}
\section{Morphological analysis}
\label{obs}

%\subsection{Roughness and correlations, Family-Vicsek scaling}
\subsection{Roughness and correlations}

The observables that we employ in the analyses of the data resulting from our simulations have been first proposed in the context of surface kinetic roughening \cite{barabasi,krug97,halpinhealy,takeuchi}. In such settings, they are applied on a height field, $h({\bf x},t)$, giving the local height of an interface growing above point ${\bf x}\in \mathbb{R}^d$ on a $d$-dimensional substrate. In the case of synchronization, the direct application of analogous observables on another scalar field, namely the phase field, $\phi({\bf x},t)$, where ${\bf x}$ is now the position of an oscillator fixed in space, leads to revealing results on the synchronization process, as recently shown in Refs.~\cite{PRR1,PRR2,PRE}.

The fluctuations of the local phases around the mean value are captured by the global width or roughness \cite{barabasi,halpinhealy,krug97}
\begin{equation}
W(L,t) \equiv \langle \overline{[\phi({\bf x},t)-\overline{\phi}(t)]^2} \rangle^{1/2},
\label{w}
\end{equation}
where the overbar denotes a spatial average in a system of linear (substrate) size $L$ and the angular brackets denote averaging over different randomness realizations.  Our interest lies in couplings sufficiently strong so saturation of the roughness may be eventually attained. The key point is that differences between oscillator phases that do not evolve at the same effective frequency $\omega^{\text{eff}}_i$, Eq.\ (\ref{omegaeff}), must grow steadily in time for long times. For this reason, the saturation of $W(L,t)$ [or equivalently, that of the correlations in Fourier $S({\bf k},t)$ or real space $G({\bf r},t)$, see the definitions below] as $t\to\infty$, which shows that the phase differences stop growing at some time, indicates the presence of synchronization in the sense of frequency locking mentioned above.

Critical dynamics in surface kinetic roughening implies that surface height values are statistically correlated for distances smaller than a correlation length $\xi(t)$ which increases with time as a power law, $\xi(t) \sim t^{1/z}$, where $z$ is the dynamic exponent. The same correlations are at play here between the phases of the oscillators distributed across space. Such an increase takes place until $\xi(t)$ reaches a value comparable to $L$, which results into the width saturating at a steady-state, size-dependent value $W(L,t\gg L^z) \sim L^\alpha$. Here, $\alpha$ is the roughness exponent, which is related with the fractal dimension of the interface profile $\phi(\mathbf{x},t)$ \cite{Mozo22}. In a wide variety of physical contexts and conditions, the global roughness satisfies the Family-Vicsek (FV) scaling Ansatz
\cite{barabasi,halpinhealy,krug97,vicsek}
\begin{equation}
W(L,t) = t^{\beta} f(L/\xi(t)),
\label{fv}
\end{equation}
where the scaling function $f(y) \sim y^\alpha$ for $y\ll 1$, while $f(y)$ reaches a constant value for $y\gg 1$. The ratio $\beta = \alpha/z$ is known as the growth exponent, and characterizes the short-time behavior of the roughness, $W(t)\sim t^{\beta}$. Notably, the FV Ansatz is verified by classical models of equilibrium critical dynamics \cite{tauber14}. Away from equilibrium, it is also verified by representatives of important universality classes of kinetic roughening, like those of the KPZ and EW equations, which are characterized by the set of $(\alpha,z)$ exponent values and their dependence on the substrate dimension $d$ \cite{barabasi,halpinhealy,krug97}. Lattices of both phase and limit-cycle oscillators subjected to time-dependent noise have been recently shown to be in such universality classes, hence to display FV scaling \cite{PRE}.

Beyond global quantities like $W(L,t)$, the GSI occurring in kinetic roughening systems also impacts the behavior of correlation functions. A particularly useful one is the height-difference correlation function,
\begin{equation}
    G({\bf r},t) \equiv \langle \overline{[\phi({\bf x}+{\bf r},t) - \phi({\bf x},t)]^2} \rangle.
\label{grt}
\end{equation}
In our cases of interest, due to rotational invariance, the correlations only depend on $\ell \equiv |{\bf r}|$. For FV scaling %with $\alpha < 1$ (which is the standard, self-affine scaling),
$G(\ell,t)$ scales differently depending on how $\ell$ compares with the correlation length \cite{barabasi,halpinhealy,krug97},
\begin{equation}
    G(\ell,t) \sim \left\{
    \begin{array}{lr}
        t^{2 \beta},& \text{if } t^{1/z} \ll \ell, \\
        \ell^{2 \alpha}, & \text{if } \ell  \ll t^{1/z}.
    \end{array}
    \right\} = \ell^{2\alpha} g(\ell/\xi(t)) ,
\label{glt}
\end{equation}
where we are assuming $\ell < L$ and $g(y)$ is a suitable scaling function \cite{barabasi,krug97}. Again, this behavior has been seen in oscillator lattices in the presence of time-dependent noise \cite{PRE}. Note generally that $G(\ell,t)$ scales like the square of the (local) width $w(\ell,t)$ restricted to a region of linear size $\ell$, namely, for $w(\ell,t) \equiv \langle \overline{[\phi({\bf x},t)-\overline{\phi}]^2} \rangle^{1/2}$ (the spatial average is here restricted to such a region), $G(\ell,t) \sim w^2(\ell,t)$  
% One can see that the scaling form in Eq.\ (\ref{glt}) only reflects the growth and saturation discussed above for the whole system, but now restricted to a region of linear size $\ell$ 
\cite{lopezphysa}. 
%A related correlation function that is also frequently studied \cite{takeuchi} is the so-called {\it height covariance}
%\begin{equation}
%C(\mathbf{r},t) \equiv \langle \overline{\phi(\mathbf{x},t)  \phi(\mathbf{x}+\mathbf{r},t)} \rangle - \langle \bar{\phi}(t) \rangle^2 ,
%    \label{eq:cov}
%\end{equation}
%such that, again under the assumption of rotational invariance, $G(\ell,t)= 2 [W^2(t)-%C(\ell,t)]$ \cite{krug97}.

%Self-affinity is an independent additional condition to Family-Vicsek scaling, and in fact there are growth process that are neither self-affine nor display Family-Vicsek scaling \cite{lopez}.

Whenever the roughness exponent $\alpha \geq 1$ (see the next subsection), as in the case of oscillators in the presence of columnar disorder \cite{PRR1,PRR2}, it is useful to consider \cite{lopezphysa, siegert96} an alternative two-point correlation function, namely, the surface structure factor, i.e., the power spectral density (PSD) or structure factor of the phase fluctuations, defined as
\begin{equation}
S({\bf k},t) \equiv \langle \hat{\phi}({\bf k}, t)\hat{\phi}(-{\bf k}, t)\rangle = \langle |\hat{\phi}({\bf k}, t)|^2\rangle,
\label{Sk}
\end{equation}
where $\hat{\phi}({\bf k},t)\equiv\mathcal{F}[\phi(\mathbf{x},t)]$ is the space Fourier transform of $\phi(\mathbf{x},t)$ and $\mathbf{k}$ is $d$-dimensional wave vector. The FV Ansatz now reads
\begin{equation}
S({\bf k},t) = k^{-(2  \alpha +d)} s_{\rm FV}(k t^{1/z}) ,
\label{SkFV}
\end{equation}
with $s_{\rm FV}(y)$ approaching a constant value for $y\gg 1$ and $s_{\rm FV}(y) \propto y^{2\alpha +d}$ for $y\ll 1$. This has again be observed for oscillator lattices subjected to time-dependent noise \cite{PRE} and also for lattices with columnar disorder in the particular case of Kuramoto coupling (i.e.\! KS coupling with $\delta = 0$) \cite{PRR1}. This scaling form can be derived by realizing that $W^2(L,t)$ equals the integral of $S({\bf k},t)$ over wave vector space (Parseval's theorem) \cite{barabasi}. Likewise, $S(\mathbf{k},t)$ is analytically related with, e.g., $G(\mathbf{r},t)$ via space Fourier transforms \cite{krug97}.

\subsection{Anomalous scaling}

The oscillator lattices with columnar disorder studied in Refs.~\cite{PRR1,PRR2} actually display a related but different (i.e.\! non-FV) form of GSI behavior termed anomalous scaling \cite{schroeder93,dassarma,krug97,lopez,ramasco,cuerno04}. While for standard FV systems height fluctuations at local distances $\ell \ll L$ scale with the same roughness exponent as global fluctuations do at distances comparable with the system size $L$, in systems displaying anomalous scaling local and global fluctuations scale with different roughness exponents, i.e. $w(\ell,t\gg \ell^z) \sim \ell^{\alpha_\text{loc}}$ with $\alpha_\text{loc} \neq \alpha$. The anomalous scaling that occurs in the synchronization process is most conveniently identified by means of the structure factor, as a new independent exponent $\alpha_s$ (termed spectral roughness exponent) appears in the dominant contribution in Fourier space, namely \cite{ramasco},
\begin{equation}
S(k,t) = k^{-(2  \alpha +d)} s(k t^{1/z}) ,
\label{Skan}
\end{equation}
where $s(y) \propto y^{2(\alpha - \alpha_s)}$ for $y\gg 1$ and $s(y) \propto y^{2\alpha +d}$ for $y\ll 1$.
Equation \eqref{Skan} generalizes the FV Ansatz, Eq.\ \eqref{SkFV}, which is retrieved if $\alpha_s=\alpha$, only found for columnar disorder in the particular case of K coupling thus far \cite{PRR1}. Generally speaking, the exponent $\alpha_s$ and its value relative to that of $\alpha$ determine the type of anomalous scaling \cite{ramasco}.

For generic coupling functions, whenever the columnar disorder dominates, we have $\alpha_s>1$, for which the correlation function scales as \cite{lopezphysa}
\begin{equation}
    G(\ell,t) \sim \left\{
    \begin{array}{lr}
        t^{2 \beta},& \text{if } t^{1/z} \ll \ell \ll L, \\
        \ell^{2 \alpha_\text{loc}} t^{2(\alpha-\alpha_\text{loc})/z}, & \text{if } \ell  \ll t^{1/z}  \ll L.
    \end{array}
\right.
\label{glt_mod}
\end{equation}
This means that the two-point correlations keep increasing (anomalously) with time even at distances which are smaller than the correlation length (which would saturate in case of FV scaling); in contrast, now two-point correlations only saturate at $\ell^{2 \alpha_\text{loc}} L^{2(\alpha-\alpha_\text{loc})}$ when $t^{1/z} \sim L$. If $\alpha = \alpha_s >1$, as for K coupling with columnar disorder, the anomalous scaling is termed super-rough \cite{lopez}, due to the large interface fluctuations that occur. The dynamic scaling ansatz satisfied by the structure factor is FV in this case, Eq.~(\ref{SkFV}), but $\alpha_\text{loc} = 1 \neq \alpha$. Otherwise, if $\alpha \neq \alpha_s$ with both exponents being larger than 1, faceted anomalous scaling takes place \cite{ramasco}, and again $\alpha_\text{loc} = 1$. There exist cases (like that of the tensionless KPZ equation \cite{rodriguez-fernandez22}) in which $\alpha_s<1$, leading to so-called intrinsic anomalous scaling for which $\alpha_\text{loc}$ need not equal 1 \cite{lopez}. But in the synchronization process it is faceted scaling that is generically observed whenever columnar disorder is the dominating type of randomness \cite{PRR1,PRR2}.

%\subsection{Universal fluctuations in the synchronization process}
\subsection{Universal fluctuations}

For rough surfaces, another observable of interest, which has received much attention during the last decade, is the probability distribution function (PDF) of the fluctuations of the heights $h({\bf x},t)$ [in our case, that of the phases $\phi({\bf x},t)$], around their mean.
%average power-law growth $t^\beta$.
By a straightforward adaptation of the definition employed in the kinetic roughening literature \cite{kriecherbauer10,halpinhealy,takeuchi}, we focus on the PDF of
\begin{equation}
\varphi_i\equiv \frac{\delta \phi_i(t_0+\Delta t) - \delta \phi_i(t_0)}{(\Delta t)^\beta} ,
\label{fluct}
\end{equation}
where $\delta \phi_i(t) = \phi_i(t) - \overline{\phi}(t)$, $t_0$ is a reference time beyond the initial transient dynamics, and $t_0 +\Delta t$ is some intermediate time within the growth regime for which $W(L,t) \sim t^{\beta}$. The division by $(\Delta t)^\beta$ removes the systematic increase of the fluctuations in time so that, remarkably, the PDF of $\varphi_i$ reaches a universal, time-independent form \cite{kriecherbauer10,takeuchi}.
Important examples in the kinetic roughening literature are, e.g., the Gaussian distribution for the linear EW equation \cite{barabasi,krug97} and a TW PDF (whose precise form depends, e.g., on boundary conditions) for the KPZ equation \cite{kriecherbauer10,takeuchi}, both of them with time-dependent noise.
%average contribution to the growth and allows us to focus on the distribution of fluctuations around it.
%Recently, the ubiquitous TW PDFs \cite{makey20} have been shown to accurately describe the distribution of (similarly defined) fluctuations of models in the KPZ universality class \cite{kriecherbauer10,takeuchi}.

Except for K coupling, which is associated with Gaussian fluctuations, all results contained in Refs.~\cite{PRR1,PRR2,PRE} for generic coupling functions which do not possess odd symmetry show TW fluctuations associated with the Gaussian Orthogonal Ensemble (GOE) of random matrix theory, as expected for the 1D KPZ equation with PBCs \cite{kriecherbauer10,takeuchi}. In the case of limit-cycle oscillators, the absence of such a symmetry from the functions giving the coupling between phases in their phase-reduced approximations \cite{pietras} has been confirmed in some cases to the lowest order of approximation \cite{PRR2}, and it is expected to hold quite generally, even more so if higher-order terms are considered.

\section{Odd symmetry of the coupling function and the critical dynamics of synchronization}

In this section, we investigate the influence of the odd symmetry of the coupling function on the GSI behavior characterizing the synchronization process. To this end we study the observables discussed in the previous section based on data from simulations for the five coupling functions considered in Sec.~\ref{numsim}. Of these, KH1 and KH2 are the main novelty of the present analysis, as they allow us to explore large-scale dynamical features that they share with the other odd-symmetric function K, while a similar analysis in the case of K and KS was already included in Refs.~\cite{PRR1,PRE}, shown here for comparison. CH is considered to better illustrate the scaling for non-odd functions, as that coupling function is not affected by crossover effects that are observed for KS coupling. We first study phase oscillators in the presence of time-dependent noise and later on address the effect of columnar disorder (i.e.\! a random assignment of the natural frequencies). The simultaneous presence of these two types of randomness shall be explored in Section \ref{comp}.

\subsection{Time-dependent noise}
\label{therm}

As described in Secs.\ \ref{sec:1dmodels} and \ref{numsim} we consider one-dimensional systems of oscillators ($d=1$), which are now free from columnar disorder ($\omega_i = 0$ for all $i\in \{1,2,\ldots, L\}$, or, equivalently, $\sigma_\text{col} = 0$). The equation of motion is thus
\begin{equation}
\frac{d \phi_i(t)}{d t} = \Gamma[\phi_{i+1}(t) -\phi_i(t)] + \Gamma[\phi_{i-1}(t) -\phi_i(t)] + \xi_i(t),
\label{eq1therm}
\end{equation}
where the oscillators are at the sites of a ring, i.e.\! $\phi_{L+1}(t) \equiv \phi_1 (t)$ and $\phi_{0}(t) \equiv \phi_L (t)$ (PBCs). The coupling function $\Gamma[\Delta \phi(t)]$ will take the five different forms listed in Section \ref{numsim}, including some that do (K, KH1, KH2) or do not (KS,CH) satisfy the odd symmetry $\Gamma[\Delta \phi(t)]+\Gamma[-\Delta \phi(t)] = 0$.

\begin{figure}[t!]
\includegraphics[scale=0.33]{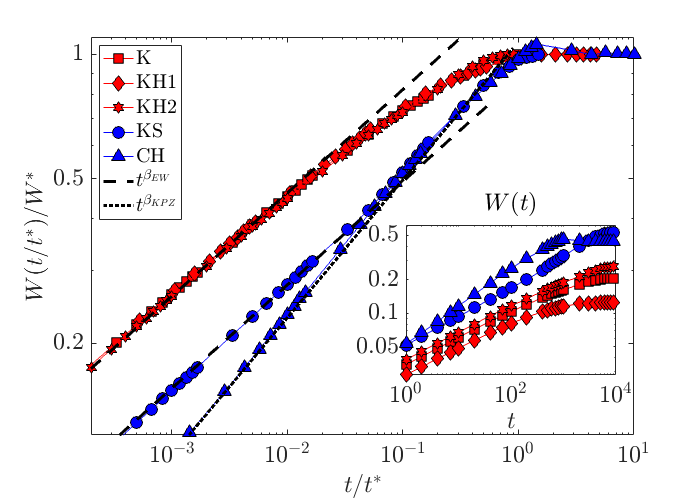}
\caption{Roughness $W(L,t)$ with (main panel) and without (inset) normalization by the saturation value $W^*$ and time $t^*$ for a ring of $L=1000$ phase oscillators affected by time-dependent noise. Different symbols represent different coupling functions (see legend), with all symbols in red corresponding to odd-symmetric coupling functions (K, KH1, KH2) and those in blue to non-odd-symmetric coupling functions (KS, CH). Power-law scalings $W \sim t^{\beta}$ with the EW and KPZ growth exponents are used as a guide to the eye (see legend). Results based on 10000 realizations.}\label{figThW}
\end{figure}

For such systems, the first observable of those discussed in Section \ref{obs} that we consider is the roughness $W(L,t)$, see Eq.~(\ref{w}). This is displayed in the inset of Fig.~\ref{figThW} for all five coupling functions under investigation with fixed $L = 1000$ [hence the notation $W(t)$]. In the main panel of Fig.~\ref{figThW}, $W(t)$ appears normalized by the saturation value $W^* = \lim_{t\to \infty} W(t)$ as a function of the time variable normalized by the saturation time $t^*$, which is defined as the smallest time for which the saturation value $W^*$ is reached. The normalized form $W(t/t^*)/W^*$ provides the clearest view on the large-scale behavior of the system, as it is not affected by non-universal (system-dependent) saturation values like $W^*$ and $t^*$. It clearly shows that all couplings with the odd symmetry of K, i.e.\! KH1 and KH2 as well, have a roughness that grows in time in a power-law fashion with the EW growth exponent,  $W(t)\sim t^{\beta_\text{EW}}$ for $\beta_\text{EW} = 1/4$. A similar behavior is obtained for KS, but only at the initial stages of the growth. Then the system crosses over to a KPZ scaling $W(t)\sim t^{\beta_\text{KPZ}}$ for $\beta_\text{KPZ} = 1/3$ and continues like that until saturation. By contrast, KPZ scaling is observed throughout the growth regime for the case of CH coupling.  

The type of crossover observed for KS coupling is known to occur in the KPZ equation \cite{forrest}, yet it was absent from our previously reported results on the synchronization of oscillators affected by time-dependent noise \cite{PRE}. In that reference, the same noise strength $\sigma_\text{tdep} = 0.1$ was employed, yet the coupling strength of the KS function was only $K=1$ instead of $K=20$, which may account for this difference. The latter coupling strength was chosen here so that the same value can be employed in the columnar-disorder case, for which $K=1$ does not suffice to reach synchronization (i.e.\! saturation). It is likely that the absence of such crossover from the CH coupling case is due to the dominance of cosine terms for that coupling function, which corresponds to a stronger KPZ nonlinearity in the continuum approximation, see Sec.~\ref{sec:contapp}.

\begin{figure}[t!]%[h!]
\includegraphics[scale=0.33]{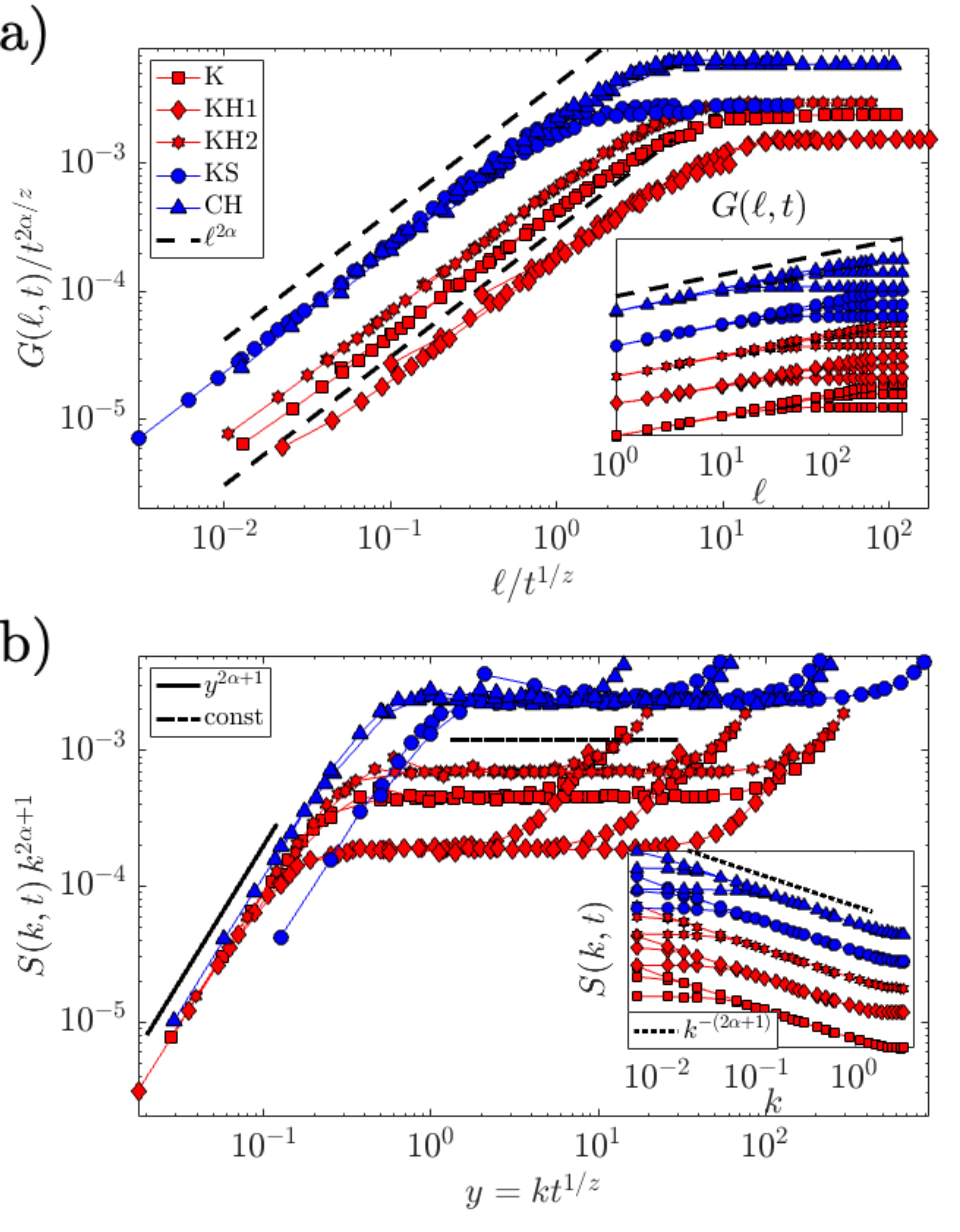}
%\vspace{-0.8cm}
\caption{Correlations $G(\ell,t)$ (a) and structure factor $S(k,t)$ (b) for a ring of $L=1000$ phase oscillators affected by time-dependent noise. Different symbols represent different coupling functions (see legend), with all symbols in red corresponding to odd-symmetric coupling functions (K, KH1, KH2) and those in blue to non-odd-symmetric coupling functions (KS and CH). Original data for $G(\ell,t)$ and $S(k,t)$ are displayed in the insets for times $t/t^* = 1/16^z, 1/4^z$ and $1$ (with an arbitrary vertical displacement for ease of visualization), while their FV rescaled forms are shown in the main panels. For odd-symmetric coupling functions $z=z_\text{EW} = 2$ and for KS and CH coupling $z = z_\text{KPZ} = 3/2$, while in both cases $\alpha = \alpha_\text{EW} = \alpha_\text{KPZ} = 1/2$. Results based on 10000 realizations.}\label{figThCorrSk}
\end{figure}

We next consider the correlations $G(\ell,t)$, defined as in Eq.~(\ref{grt}), which for FV scaling are expected to scale as in Eq.~(\ref{glt}). In  Fig.~\ref{figThCorrSk} (a) the correlations are shown with (main panel) and without (inset) FV rescaling. For each coupling function under consideration $G(\ell,t)$ is shown at times corresponding to $t/t^* = 1/16^z, 1/4^z$ and $1$. Here, $z = z_\text{EW}=2$ for K, KH1, and KH2, and $z = z_\text{KPZ} = 3/2$ for KS and CH, according to the universal GSI behavior expected for sytems with and without odd symmetry in the coupling function, respectively. At these time points the normalized correlation length $\xi(t)/L$ reaches (roughly speaking, as prefactors are not included) the values $1/16, 1/4$, and $1$. In the inset, the curves without rescaling (which have been subjected to an arbitrary vertical displacement for ease of visualization) show how the correlation length $\xi(t)$ increases and eventually spans the whole system upon saturation, with $G(\ell,t)$ for $\ell < \xi(t)$ reaching a steady-state value that is not modified thereafter. The rescaling following Eq.~(\ref{glt}) in the main panel shows an excellent collapse of $G(\ell,t)$ for all three instants of time (and many others that have been inspected, not shown). As the roughness exponent values coincide, $\alpha_\text{EW} = \alpha_\text{KPZ} = 1/2$, the power-law behavior $G(\ell,t) \sim \ell^{2\alpha}$ for short length scales compared to $\xi(t)$ is the same for all coupling functions considered, yet the value of $z$ in the rescaling is chosen according to the GSI behavior (i.e.\! $z_\text{EW}$ for K, KH1 and KH2, $z_\text{KPZ}$ for KS and CH).

The PSD $S(k,t)$, defined as in Eq.~(\ref{Sk}), is illustrated in Fig.~\ref{figThCorrSk} (b) with (main panel) and without (inset) FV rescaling (\ref{SkFV}). The same instants of time used for displaying the correlations in Fig.~\ref{figThCorrSk} (a) are employed here for each coupling function, again with the non-rescaled values being vertically shifted for visualization purposes. The collapse is excellent for all couplings, and the same comments made about the critical exponents used in the case of $G(\ell,t)$ apply here as well.

An aspect related to correlations at small distances, specifically the local slopes at a distance $\ell =1$, here given by $\phi_{i+1}(t)- \phi_i(t)$ for $i=1,2,\ldots, L$, has been pointed out to be relevant in an etching model in the KPZ universality class, so much so that the macroscopic coupling parameters $\nu$ and $\lambda$ appear to be closely related to the slope distribution \cite{gomes}. We have checked that the slope distribution stabilizes very early on in our simulations with time-dependent noise for all coupling functions (not shown), and remains practically unchanged  throughout the growth regime, as in the etching model, and quite unlike what happens for columnar noise leading to anomalous scaling discussed in the next subsection, see Ref.~\cite{PRR1}. An intringuing open question is whether the parameters in the continuum approximation of synchronizing oscillators in Sec.~\ref{sec:contapp} can be related to the moments of the slope distribution.

\begin{figure}[h!]
\includegraphics[scale=0.3]{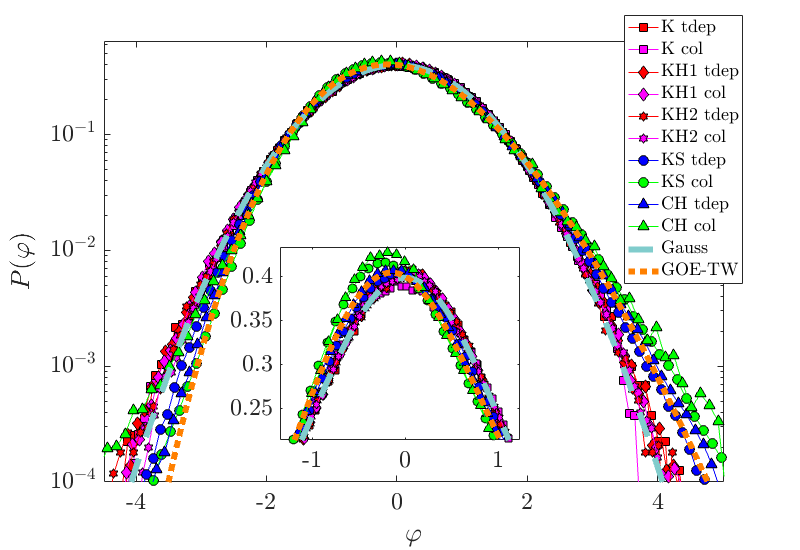}
\vspace{-0.6cm}
\caption{Histogram of fluctuations around the average growth $\varphi_i$, see Eq.~(\ref{fluct}), for a ring of $L=1000$ phase oscillators affected by time-dependent noise (red and blue symbols) or by columnar disorder (magenta and green symbols) for all five coupling functions. In each case the fluctuations correspond to (at least 9) time points in the $W \sim t^{\beta}$ growth regime (once transient crossovers have been left behind), with $\beta = \beta_\text{EW}$ ($\beta_\text{CEW}$) for K, KH1, and KH2, and $\beta = \beta_\text{KPZ}$ ($\beta_\text{CKPZ}$) for KS and CH, with time-dependent noise (columnar disorder). Different symbols represent different coupling functions (see legend), with all symbols in red (magenta) corresponding to odd-symmetric coupling functions K, KH1 and KH2, and those in blue (green) to non-odd-symmetric coupling functions KS and CH, for time-dependent noise (columnar disorder). Results based on 10000 realizations.}\label{figfluct}
\end{figure}

To conclude our investigation of oscillators affected by time-dependent noise alone, we study the PDF of the fluctuations $\varphi_i$. These are shown, normalized to zero mean and unit variance, in Fig.~\ref{figfluct} for all coupling functions, where we haven taken $\beta = \beta_\text{EW} = 1/4$ for the odd-symmetric ones and $\beta = \beta_\text{KPZ} = 1/3$ for KS and CH in the definition (\ref{fluct}). The fluctuations for the different time points in the growth regime appear to be distributed according to the same (time-independent) PDF form in each case, just as expected \cite{PRR1,PRR2,PRE}. In particular, we find that odd coupling functions have histograms that follow quite neatly the Gaussian PDF, while the others closely resemble the GOE-TW PDF, especially for CH coupling and for KS coupling in the right tail. Some deviation from the GOE-TW PDF is observed in the left tail in the case of KS coupling, which may be due to a system-size effect or insufficient statistics due to the shortness of the time interval over which KPZ behavior is observed, see again Fig.~\ref{figThW}. The GOE-TW characterizes the 1D KPZ universality class in the nonlinear growth regime under standard PBCs \cite{takeuchi}, and is the one generically found for rings of oscillators under time-dependent noise in the absence of the odd symmetry of the coupling function \cite{PRE}. Note that Fig.~\ref{figfluct} also contains results for columnar disorder, to be discussed in the next subsection.

\subsection{Columnar disorder}
\label{col}

The case of pure columnar disorder is also a one-dimensional particularization of the general model in Eq.~(\ref{eq1}), now without time-dependent noise ($\sigma_\text{tdep} = 0$). The equation of motion is thus
\begin{equation}
\frac{d \phi_i(t)}{d t} = \omega_i + \Gamma[\phi_{i+1}(t) -\phi_i(t)] + \Gamma[\phi_{i-1}(t) -\phi_i(t)].
\label{eq1col}
\end{equation}
Again the coupling function $\Gamma[\Delta \phi(t)]$ may take each of the five different forms described in Sec.~\ref{numsim}.

\begin{figure}[h!]%[h!]
\includegraphics[scale=0.34]{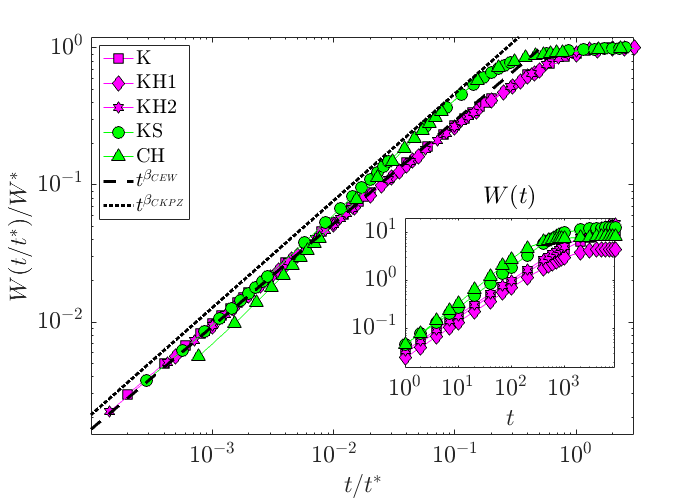}
\caption{Roughness $W(L,t)$ with (main panel) and without (inset) normalization by the saturation value $W^*$ and time $t^*$ for a ring of $L=1000$ phase oscillators affected by columnar disorder. Different symbols represent different coupling functions (see legend), with all symbols in magenta corresponding to odd-symmetric coupling functions (K, KH1 and KH2) and those in green to non-odd-symmetric coupling functions (KS and CH). Power-law scalings $W \sim t^{\beta}$ with the columar EW (CEW) and columnar KPZ (CKPZ) growth exponents are used as guides to the eye (see legend). Results based on 10000 realizations.}\label{figColW}
\end{figure}

The analysis is based on the same observables discussed in Sec.~\ref{obs} and already employed in the previous subsection for the case of time-dependent noise. We again start with the roughness, Eq.~(\ref{w}), which is shown in  Fig.~\ref{figColW} for all five coupling functions under investigation with fixed $L = 1000$. In the main panel, $W(t)$ again appears normalized by the saturation value $W^*$ and as a function of the time variable normalized by the saturation time $t^*$. The normalized form $W(t/t^*)/W^*$ clearly shows that all couplings with the odd symmetry of K have a roughness that grows in time as a power law with the columnar EW (Larkin) growth exponent, $W(t)\sim t^{\beta_\text{CEW}}$ with $\beta_\text{CEW} = 3/4$. A similar behavior is obtained for KS, but only at the very initial stages of growth. Then the system crosses over to a columnar KPZ scaling $W(t)\sim t^{\beta_\text{CKPZ}}$ for $\beta_\text{CKPZ} = 1.07/1.37 \approx 0.78$. This exponent value, which is known to be affected by non-universal corrections related to the columnar disorder distribution \cite{krughh}, has been chosen after Ref.~\cite{PRR2}. In that reference and Ref.~\cite{PRR1}, such a crossover was not observed, not even for KS coupling, possibly due to the longer times employed, $t/t^* \gtrsim 10^{-3}$, and/or to the different choice of parameters, including the support of the columnar disorder distribution, which was compact (uniform distribution), rather than extended (Gaussian distribution) as in our present simulations, see Ref.\ \cite{krug97}. The similarity between $\beta_\text{CEW}$ and $\beta_\text{CKPZ}$ (cf.\! the much larger difference found for the time-dependent scenario discussed in the previous subsection) may have also hampered observation of this crossover. For CH coupling, on the other hand, KPZ scaling is observed also for later stages of growth, as is the case for KS coupling, while the initial stages seem to be governed by an even steeper growth, whose origin is at present unclear.
%\textcolor{red}{Se sabe algo de crossovers en el caso de desorden columnar?}

\begin{figure}[h!]
\includegraphics[scale=0.33]{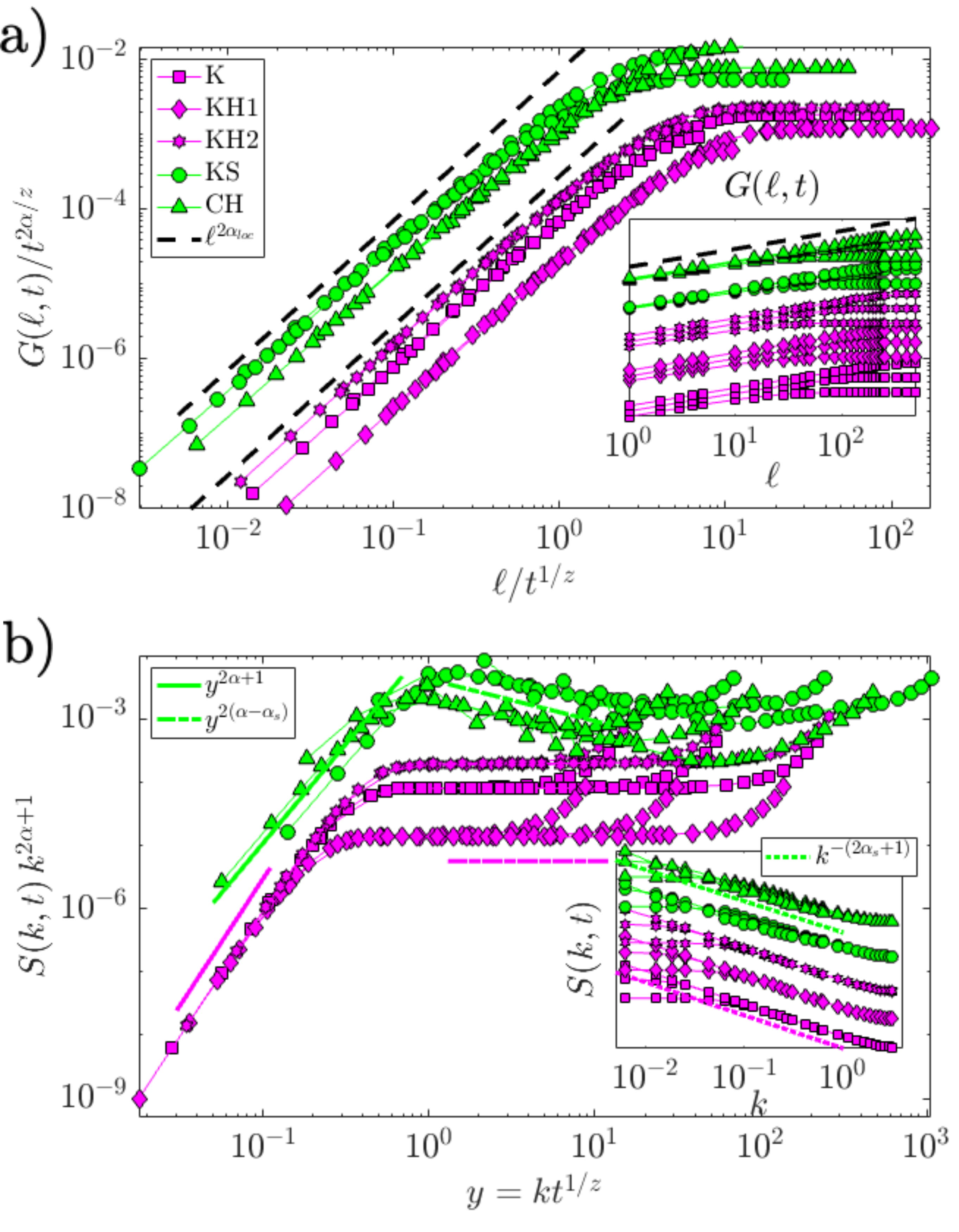}
\caption{Correlations $G(\ell,t)$ (a) and structure factor $S(k,t)$ (b) for a ring of $L=1000$ phase oscillators affected by columnar disorder at times $t/t^* = 1/16^z, 1/4^z$, and $1$. Different symbols represent different coupling functions (see legend), with all symbols in magenta corresponding to odd-symmetric coupling functions (K, KH1 and KH2) and those in green to non-odd-symmetric coupling functions (KS and CH). Unscaled data are displayed in the insets (with an arbitrary vertical displacement for ease of visualization), while their anomalous rescaled forms are shown in the main panels. For odd-symmetric coupling functions $\alpha = \alpha_s = \alpha_\text{CEW} = 3/2$ and $z=z_\text{CEW} = 2$, while for KS and CH coupling $\alpha = \alpha_\text{CKPZ} = 1.07$, $\alpha_s = 1.40$, $z = z_\text{CKPZ} = 1.37$; in both cases we take $\alpha_\text{loc} = 1$. Results based on 10000 realizations.}\label{figColCorrSk}
\end{figure}

A clearer view on the scaling forms at play is obtained from the independent measurement of the roughness $\alpha$ and dynamic $z$ exponents, whose ratio is the growth exponent $\beta=\alpha/z$; hence, we next turn to correlation functions. Note that, in the presence of columnar disorder, anomalous scaling is expected \cite{PRR1,PRR2}, so that $G(\ell,t)$ should now scale as in Eq.~(\ref{glt_mod}). In Fig.~\ref{figColCorrSk} (a) the correlations are shown with (main panel) and without (inset) such rescaling. For each coupling function under consideration, $G(\ell,t)$ is shown again at times such that  $t/t^* = 1/16^z, 1/4^z$, and $1$. We take the Larkin model value $z = z_\text{CEW}=2$ for K, KH1, and KH2, and $z = z_\text{CKPZ} = 1.37$ for KS and CH, in agreement with results reported in Ref.~\cite{PRR2}. The unscaled curves (which have been subjected to an arbitrary vertical displacement for ease of visualization) show how the correlation length $\xi(t)$ increases and eventually spans the whole system, yet $G(\ell,t)$ for $\ell < \xi(t)$ keeps on increasing until global saturation when $\xi(t)\approx L$, which is one of the hallmarks of anomalous scaling. The rescaling following Eq.~(\ref{glt_mod}) shows an excellent collapse of $G(\ell,t)$ for all three instants of time (and many others that have been inspected, not shown). Regarding the roughness exponent, for K, KH1, and KH2 we take $\alpha_\text{CEW} = 3/2$ and for KS $\alpha_\text{CKPZ} = 1.07$, as in Refs.~\cite{PRR1,PRR2}. The power laws highlighted in the figure are of the form $\ell^{2\alpha_\text{loc}}$ for which we have taken $\alpha_\text{loc} = 1$. This is the theoretical value of the columnar EW equation, characterized by super-rough scaling. It is also extremely close to the value of 0.96 or 0.97 (the actual discrepancy being possibly a finite-size effect) numerically found in Refs.~\cite{PRR1,PRR2} for generic (non-odd) couplings leading, as is the case for KS and CH here, to the full set of columnar KPZ exponent values at large space-time scales.

The structure factor $S(k,t)$ is illustrated in Fig.~\ref{figColCorrSk} (b) with (main panel) and without (inset) rescaling as in Eq.\ (\ref{Skan}). The same instants of time used for displaying the correlations in Fig.~\ref{figColCorrSk} (a) are employed here for each coupling function, again with the non-rescaled values being vertically shifted for visualization purposes. For K, KH1, and KH2 the spectral exponent $\alpha_s = \alpha_\text{CEW} = 3/2$, while for KS and CH $\alpha_s = 1.40$, as in Refs.~\cite{PRR1,PRR2}. The collapse is excellent in all cases, and the same comments made about the critical exponents used in the collapse of $G(\ell,t)$ apply here as well. The power laws characterizing the scaling function in Eq.~(\ref{Skan}) here depend on the GSI behavior (columnar EW or columnar KPZ), which is why they are differentiated by the color in the figure (magenta for odd couplings, green for non-odd couplings), as the exponents involved are different in each case. The most conspicuous difference is the nonzero value of $\alpha - \alpha_s$, used as an exponent of the rescaled variable, $y^{2(\alpha-\alpha_s)}$, for non-odd KS and CH couplings, which is a feature of faceted scaling observed in the columnar KPZ equation \cite{szendro}. This is absent from the scaling function for the odd couplings leading to columnar EW GSI \cite{purrello}, characterized by super-rough scaling with $\alpha_s = \alpha$.

To conclude our investigation of oscillators affected by columnar disorder alone, we study the PDF of the fluctuations, which are shown, normalized to zero mean and unit variance, in Fig.~\ref{figfluct} for all coupling functions under consideration, where $\beta = \beta_\text{CEW}$ for the odd-symmetric ones (K, KH1 and KH2) and $\beta = \beta_\text{CKPZ}$ for non-odd functions (KS and CH). As in the time-dependent-noise case, we find that odd coupling functions have histograms that follow quite neatly the Gaussian PDF, while the others closely resemble the GOE-TW PDF, which is the one generically found for rings of oscillators under columnar disorder in the absence of the odd symmetry of the coupling function \cite{PRR1, PRR2}. There is a deviation at the peak of the distribution (see inset), which may be due to insufficient statistics, yet both tails of the PDF are followed quite closely by the data down to very small probabilities, with the sole exception of CH coupling on the left tail. Except for such minor discrepancies, the same PDFs of the fluctuations around the average growth, Eq.\ (\ref{fluct}), including their dependence on the odd symmetry of the coupling function, are found for either type of randomness considered, whether it is time-dependent noise or columnar disorder, which is remarkable considering how different the respective scalings are. In the next section we investigate the more complex situation in which both types of randomness simultaneously concur.

\section{Competing forms of randomness:  time-dependent noise vs columnar disorder}
\label{comp}

In this section we consider a one-dimensional version of the model in Eq.~(\ref{eq1}) with both time-dependent noise and columnar disorder. The equation of motion is thus
\begin{equation}
\frac{d \phi_i(t)}{d t} = \omega_i + \Gamma[\phi_{i+1}(t) -\phi_i(t)] + \Gamma[\phi_{i-1}(t) -\phi_i(t)]+ \xi_i(t).
\label{eq1comp}
\end{equation}
Our objective is to study the impact of the competition between the two forms of randomness on the emerging GSI behavior, for which we will vary the strengths of the time-dependent noise and the columnar disorder in our simulations. In particular, we will consider:
\begin{itemize}
\item $\sigma_\text{col} = 0.1$ and $\sigma_\text{tdep} = 0.01$, which we term columnar-disorder dominated (CDD);
\item $\sigma_\text{col} = 0.01/0.02$ for K/CH coupling and $\sigma_\text{tdep} = 0.1$ (in both cases), which shows a clear crossover between both forms of randomness (Cross);
\item $\sigma_\text{col} = 0.0001$ and $\sigma_\text{tdep} = 0.1$, which we term time-dependent-noise dominated (TND).
\end{itemize}
The CDD case will be shown to be essentially indistinguishable from the pure columnar-disorder situation of Sec.~\ref{col}, and the same can be said about the TND and the pure time-dependent-noise setting of Sec.~\ref{therm}. The most interesting situation will be shown to be Cross, where the competition effectively leads to a crossover from time-dependent-noise GSI at the initial stages to columnar-disorder GSI at later stages of the growth regime. The choice of standard deviations for the three cases listed above arises from the observation that, for a comparable strength, it is always the columnar disorder that prevails, and even when there is an apparent competition in the form of a crossover, it is the dominant form of randomness in the latest stages of the synchronization process. It seems that one has to reduce the columnar disorder strength by several orders of magnitude with respect to the strength of the time-dependent noise in order for the latter to dominate the GSI behavior until saturation (see the results below). The parameters are chosen to highlight the different behaviors in as clear a manner as possible, which in the case of Cross requires a small difference in the columnar-disorder strengths for K and CH coupling.

As the influence of the odd symmetry of the coupling function $\Gamma[\Delta \phi(t)]$ has been clarified in the previous section, for simplicity we limit the list of coupling functions investigated to two of those discussed in Section \ref{numsim}, including one representative example of an odd-symmetric function and one that does not possess that symmetry, namely K and CH. The analysis is again based on an exploration of the observables discussed in Sec.~\ref{obs}, and the results obtained for the other coupling functions (not shown) are very similar to those reported for these two cases.

\begin{figure}[h!]%[h!]
\includegraphics[scale=0.33]{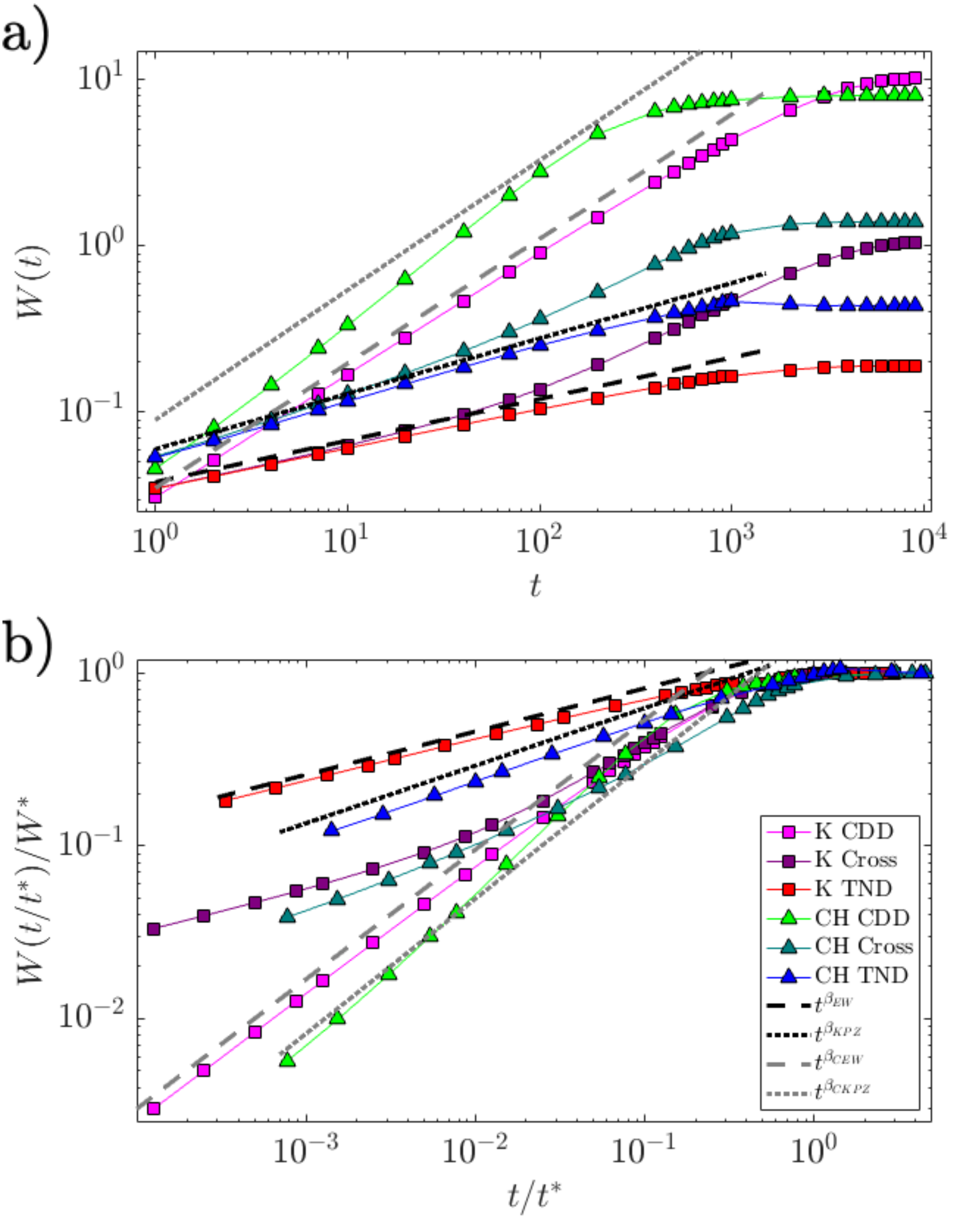}
\caption{Roughness $W(L,t)$ without (a) and with (b) normalization by the saturation value $W^*$ and time $t^*$ for a ring of $L=1000$ phase oscillators affected by time-dependent noise and columnar disorder. Different symbols represent different coupling functions (K or CH) and different colors highlight different randomness-competition configurations (CDD, Cross, or TND). Power-law scalings $W \sim t^{\beta}$ with the relevant time-dependent (EW, KPZ) and columar (CEW, CKPZ) scaling exponents are used as guides to the eye (see legend). Results based on 10000 realizations.}\label{figColThW}
\end{figure}

The roughness, Eq.~(\ref{w}), is shown in Fig.~\ref{figColThW}, without [panel (a)] and with [panel (b)] normalization by the saturation values $W^*$ and $t^*$. The randomness configurations CDD and TND do not show a significant departure from the roughnesses previously inspected in Figs.~\ref{figThW} and \ref{figColW}. Cross is more interesting in that it shows a crossover from the corresponding scaling (EW for K, KPZ for CH) for time-dependent noise at early times [which is more conveniently observed in the unnormalized $W(t)$ in Fig.~\ref{figColThW} (a)] to columnar-disorder scaling at later times [for which the normalized $W(t/t^*)/W^*$ in Fig.~\ref{figColThW} (b) provides a clearer view].

\begin{figure}[t!]
\includegraphics[scale=0.33]{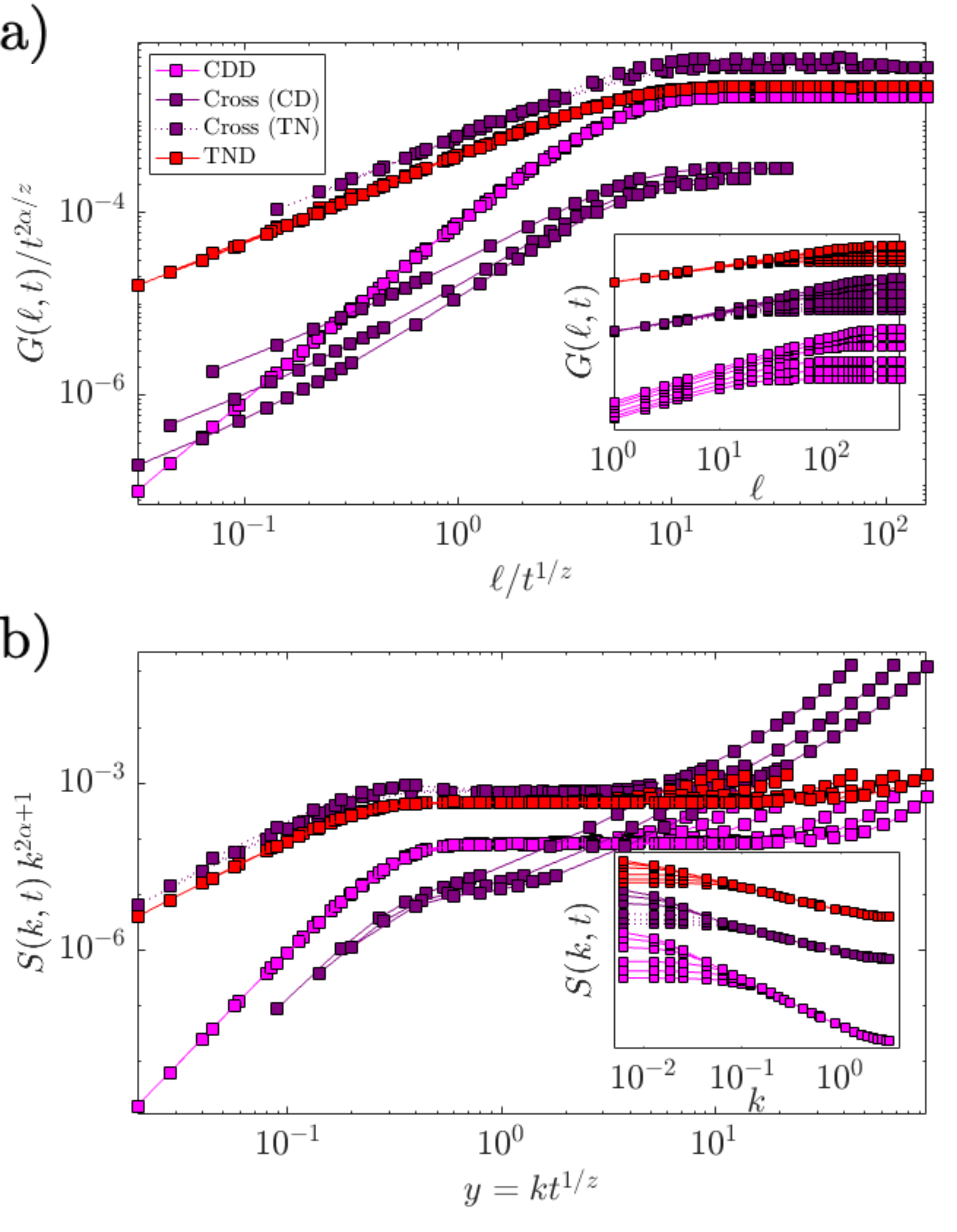}
\caption{Correlations $G(\ell,t)$ (a) and structure factors $S(k,t)$ (b) for a ring of $L=1000$ phase oscillators with K coupling, affected by time-dependent noise and columnar disorder, at times $t= 10, 20, 50, 200, 500$, and $1000$. Different symbols represent different randomness-competition configurations (see legend). Unscaled data are displayed in the insets (with an arbitrary vertical displacement for ease of visualization), while their rescaled forms are shown in the main panels (also with an arbitrary vertical displacement). For CDD $\alpha = \alpha_\text{CEW} = 3/2$ and $z=z_\text{CEW} = 2$, while for TND $\alpha = \alpha_\text{EW} = 1/2$ and $z=z_\text{EW} = 2$. In the case of Cross, for $t= 10, 20$, and $50$ rescaling has been achieved with $\alpha = \alpha_\text{EW}$ and $z=z_\text{EW}$, while for $t=200, 500$, and $1000$ we have taken $\alpha = \alpha_\text{CEW}$ and $z=z_\text{CEW}$. Results based on 10000 realizations.}\label{figColThCorrSk_K}
\end{figure}

A more detailed description of the behavior displayed by Cross is provided by the two-point correlations, where the rescaling must now be performed with the exponents corresponding to time-dependent noise at the early stages, and with those of the columnar-disorder exponents for later times reaching into saturation.  This is shown, for the case of K coupling, in Fig.~\ref{figColThCorrSk_K}, both in real space as given by the function $G(\ell,t)$, see panel (a), and in Fourier space as given by the structure factor $S(k,t)$, see panel (b).  More specifically, (both real- and Fourier-space) correlations  in Fig.~\ref{figColThCorrSk_K} are shown for $t= 10, 20, 50, 200, 500$ and $1000$ without rescaling in the insets. In the main panels, correlations for CDD and TND have been rescaled for all six time points with the exponents of the EW equation with columnar disorder and time-dependent noise, respectively. In the case of Cross, the first three time points have been rescaled using the EW exponents with time-dependent noise Cross [denoted Cross (TN) in the legend], while the last three have been rescaled using the exponents of columnar EW [denoted Cross (CD) in the legend]. Such division is motivated by the roughness results displayed in Fig.~\ref{figColThW}.  In both regimes we find that the rescaling gives a good collapse of the data for the relevant spatial scales, confirming the existence of a crossover in time from a regime dominated by time-dependent noise into one where columnar disorder prevails. In fact, when data corresponding to the last three time points are rescaled using the thermal-noise exponents, or data for the first three time points are rescaled using the columnar-disorder exponents, the curves show conspicuous systematic deviations from ``the good ones'', indicating quite clearly that such a rescaling is inappropriate at those times (not shown).

\begin{figure}[t!]
\includegraphics[scale=0.33]{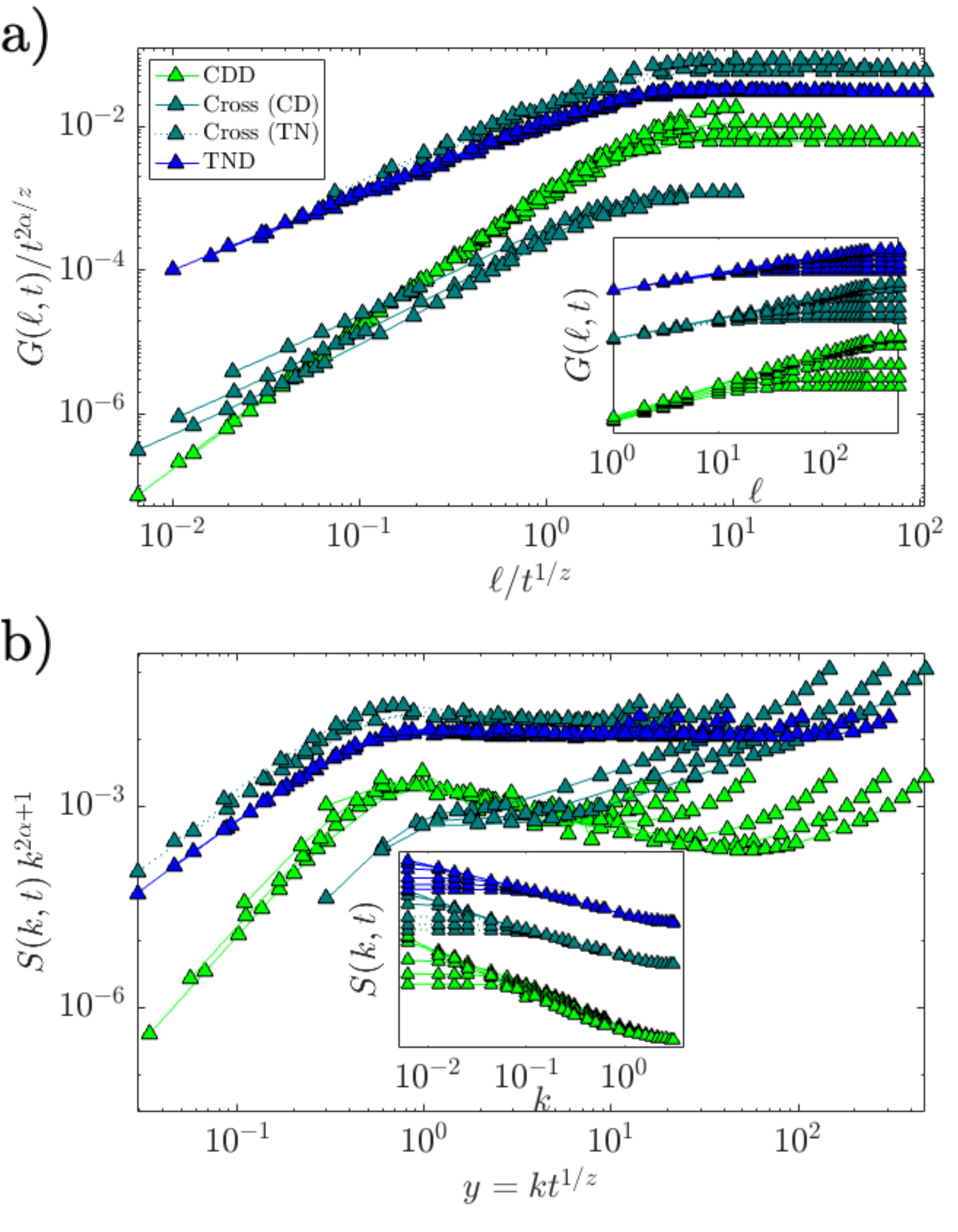}
\caption{Correlations $G(\ell,t)$ (a) and structure factors $S(k,t)$ (b) for a ring of $L=1000$ phase oscillators with CH coupling, affected by time-dependent noise and columnar disorder, at times $t= 10, 20, 50, 200, 500$, and $1000$. Different symbols represent different randomness-competition configurations (see legend). Unscaled data are displayed in the insets (with an arbitrary vertical displacement for ease of visualization), while their rescaled forms are shown in the main panels (also with an arbitrary vertical displacement). For CDD $\alpha = \alpha_\text{CKPZ} = 1.07$ and $z=z_\text{CKPZ} = 1.37$, while for TND $\alpha = \alpha_\text{KPZ} = 1/2$ and $z=z_\text{KPZ} = 3/2$. In the case of Cross, for $t= 10, 20$, and $50$ the rescaling has been achieved with $\alpha = \alpha_\text{KPZ}$ and $z=z_\text{KPZ}$, while for $t=200,500$, and $1000$ we have taken $\alpha = \alpha_\text{CKPZ}$ and $z=z_\text{CKPZ}$. Results based on 10000 realizations.}\label{figColThCorrSk_KS}
\end{figure}

For CH coupling, the correlations are given in Fig.~\ref{figColThCorrSk_KS}, again both in real space, $G(\ell,t)$, see panel (a), and in Fourier space, $S(k,t)$, see panel (b). For the same six time points considered in Fig.~\ref{figColThCorrSk_K}, we find that CDD data can be suitably rescaled by the exponents of the columnar-disorder KPZ equation, the scaling function for $S(k,t)$ showing the characteristic shape of faceted anomalous scaling \cite{ramasco}, while TND requires the exponents of the time-dependent KPZ equation. The situation for Cross is identical to that described above (now for KPZ universality instead of EW): while the first three time points show a very good rescaling using the time-dependent-noise exponents, the last three require the columnar-disorder exponents. Again, if a given form of the rescaling is employed on the data corresponding to the other regime, there is a systematic deviation and the collapse fails (not shown).

\begin{figure}[t!]%[h!]
\includegraphics[scale=0.3]{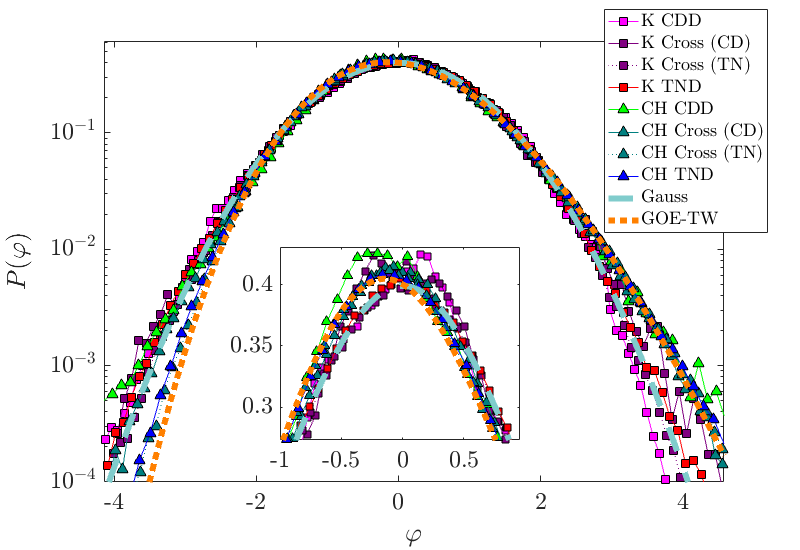}
\vspace{-0.6cm}
\caption{Histogram of phase fluctuations $\varphi_i$ around average for a ring of $L=1000$ phase oscillators affected by time-dependent noise and columnar disorder. Coupling functions K and CH, and three randomness-competition configurations, namely CDD, Cross, and TND (see text for definition) are shown. Different symbols represent different coupling functions, and different colors different levels of competition of randomness (see legend). In each case the fluctuations in Eq.~(\ref{fluct}) correspond to points within the $W \sim t^{\beta}$ growth regime (once transient crossovers have been left behind). Here, $\beta = \beta_\text{CEW}$ for K coupling, CDD, and Cross for late times dominated by the columnar disorder, CD; $\beta_\text{EW}$ for K coupling, TND, and Cross for early times dominated by the time-dependent noise, TN; $\beta = \beta_\text{CKPZ}$ for CH coupling, CDD, and Cross for late times dominated by the columnar disorder, CD; and $\beta_\text{KPZ}$ for CH coupling, TND, and Cross for early times dominated by the time-dependent noise, TN. The time windows in each case are adapted to the scaling forms observed in Fig.~\ref{figColThW}. Results based on 10000 realizations.}\label{figfluctcomp}
\end{figure}

We finally consider the PDFs of the fluctuations around the average growth, $\varphi_i$, which are displayed in Fig.~\ref{figfluctcomp}.  Both for K and CH couplings, histograms of the fluctuations are shown under randomness competitions CDD, Cross (separately for early, TN, and late, CD, times), and TND. Generally speaking, the phase fluctuations for K coupling seem to follow quite closely a Gaussian distribution, while the case of CH coupling appears non-Gaussian and closer to the GOE-TW distribution that is expected in the absence of time-dependent noise. Some significant deviations are observed on the left tail of the distribution for CH coupling in the cases of CDD and the late times (CD) of Cross, which might be related to the relatively short time window over which columnar KPZ scaling is observed in Fig.~\ref{figColThW} or with the non-universal features of the columnar-disorder exponents.

\section{Discussion and Conclusions}

Through detailed analyses based on data from numerical simulations of rings of phase oscillators with columnar disorder and time-dependent noise, we have addressed two pending questions that arise from our recent work on the large-scale dynamics of synchronizing oscillators in one dimension  \cite{PRR1,PRR2,PRE}. In those works we show that the dynamical process whereby oscillators eventually reach a synchronous state is a manifestation of nonequilibrium criticality. Our results are based on the study of observables and scaling forms previously studied in surface kinetic roughening, the reason for that being that the roughening of the ``phase interface'' on its way towards synchronization (saturation) displays universal features that have been previously understood in that context.

The first question to be addressed is one striking peculiarity previously observed when the diffusive coupling between oscillators is given by a sine function of the difference between their phases (K coupling). While other coupling functions inspected, both in the presence of columnar disorder \cite{PRR1,PRR2} and time-dependent noise \cite{PRE}, give rise to KPZ universal behavior for the corresponding type of randomness at large space-time scales, K coupling yields EW universal behavior instead \cite{PRR1,PRE}. Analytical results on the synchronization threshold \cite{ostborn, strogatz}, as well as continuum approximations \cite{PRR1}, seem to indicate that the reason for that is the odd symmetry of the coupling function. We have inspected this possibility here by explicitly simulating rings of phase oscillators with other couplings possessing the same symmetry. Our results are quite compelling in this regard and confirm that, indeed, it is the symmetry of the coupling function that leads to that distinct large-scale behavior. This further confirms the suggestion put forth in our previous work \cite{PRR2,PRE} that, for general self-sustained oscillators with an attracting limit cycle in their $d$-dimensional phase portrait (for $d\geq 2$), where such symmetry of the coupling between the phases is unlikely to hold even in the lowest orders of the phase-reduced dynamics \cite{pietras}, it is KPZ universal behavior (for the corresponding type of randomnness) that will prevail. Still EW universal traits may be important to consider in the analysis even in such cases, specially at the early stages of the synchronization process, because of the existence of crossovers in time from EW to KPZ behavior \cite{forrest}, as we have seen in Secs.~\ref{therm} and \ref{col}.

While the study of the nonequilibrium criticality of synchronization in experiments is (as far as we are aware) still to be initiated, one can reasonably anticipate that in that context both forms of randomness, i.e.\ the columnar disorder given by the different parameters of the individual oscillators and time-dependent noise (possibly of thermal origin), will compete. The influence of this competition on the large-scale GSI behavior is the second issue that we have addressed. For comparable randomness strengths it appears that columnar disorder is more dominant that time-dependent noise, in the sense that, unless the columnar disorder is made significantly weaker, at large scales one sees the universal behavior associated with that type of randomness. Yet more interestingly, when the two forms of randomness visibly manifest themselves in the evolution towards synchronization, it is a crossover from a regime dominated by time-dependent noise to one where columnar disorder prevails that is observed. In case of KPZ behavior, this crossover between the two different types of randomness can take place on top of, and compete with, the additional time crossover known to exist from EW into KPZ scaling for time-dependent noise \cite{forrest}. This may make the observation of KPZ universality quite challenging, as time-dependent EW universality might cross over to columnar-disorder universality before proper time-dependent KPZ scaling sets in.

The general qualitative conclusions that can be drawn about the impact at large scales of the competition between the two forms of randomness might, to an extent, be expected due to the persistent nature of columnar disorder as opposed to (fluctuating) time-dependent noise. Yet nontrivial questions arise in connection with this competition already in a purely kinetic roughening context \cite{krug97}. For example, whether columnar disorder is generally a relevant perturbation of the robust time-dependent noise KPZ universality class, as suggested by e.g.\ Dynamic Renormalization Group analysis of the KPZ equation with long-range correlated noise \cite{Medina89} and the relation between this system and the columnar KPZ equation \cite{Ales19}.
%Hence, it would be interesting to study e.g.\ the KPZ equation simultaneously subject to both, columnar and time-dependent noise. 
%In this regard, it is worth recalling that quenched disorder with spatial correlations... and the limit of that type of disorder for long-range correlations would correspond to columnar disorder \textcolor{red}{Rodolfo!!}

In conclusion, we provide evidence showing that, generally speaking, the universal traits of the one-dimensional KPZ equation play a crucial role in the synchronization process in one dimension. While the case of higher dimensions has not yet been studied, as far as we know, there are some indications showing that a similar analysis might be performed \cite{PRR1}; see also the phenomenological similarities between the profiles arising from the columnar KPZ equation in two dimensions \cite{szendro} and the phase interfaces displayed in Ref.~\cite{moroney}. Regarding the experimental observation of nonequilibrium criticality in the synchronization process of, e.g., nonlinear electronic circuits or chemically-oscillating media, our results on crossovers connecting different regimes (from EW to KPZ behavior for the corresponding type of randomness, and from regimes dominated by time-dependent noise to those where columnar disorder dominates) potentially provide valuable insights for the analysis of empirical data.

\section*{Acknowledgements}
This work has been partially supported by Ministerio de Ciencia e Innovaci\'on (Spain), by Agencia Estatal de Investigaci\'on (AEI, Spain, 10.13039/501100011033), and by European Regional Development Fund (ERDF, A way of making Europe) through Grants No.\ PID2021-123969NB-I00 and No.\ PID2021-128970OA-I00.

\end{document}